\documentclass[12pt,journal,draftclsnofoot,onecolumn]{IEEEtran}

\usepackage{amsfonts}
\usepackage{amssymb}
\usepackage{cite}
\usepackage[cmex10]{amsmath}
\usepackage{float}
\usepackage{color}
\usepackage{stfloats,fancyhdr}
\usepackage{mathrsfs}
\usepackage{algorithm}
\usepackage{algorithmic}
\usepackage{multirow}
\usepackage{changepage}
\usepackage[normalem]{ulem}
\usepackage{color}

\IEEEoverridecommandlockouts

\ifCLASSINFOpdf
  \usepackage[pdftex]{graphicx}
  \DeclareGraphicsExtensions{.pdf,.jpeg,.png}
\else
  \usepackage[dvips]{graphicx}
  \DeclareGraphicsExtensions{.eps}
\fi

\usepackage{subfigure}

\usepackage{fancybox,dashbox}

\begin{document}

\title{A Low-Complexity Transceiver Design in Sparse Multipath Massive MIMO Channels}

\author{Yuehua~Yu,~Peng~Wang,~He~(Henry)~Chen,~Yonghui~Li,~and Branka Vucetic
\thanks{The authors are with School of Electrical and Information Engineering, The University of Sydney, Sydney, NSW 2006, Australia (email: \{yuehua.yu, peng.wang, he.chen, yonghui.li, branka.vucetic\}@sydney.edu.au).}}


\maketitle

\begin{abstract}
In this letter, we develop a low-complexity transceiver design, referred to as semi-random beam pairing (SRBP), for sparse multipath massive MIMO channels. By exploring a sparse representation of the MIMO channel in the virtual angular domain, we generate a set of transmit-receive beam pairs in a semi-random way to support the simultaneous transmission of multiple data streams. These data streams can be easily separated at the receiver via a successive interference cancelation (SIC) technique, and the power allocation among them are optimized based on the classical waterfilling principle. The achieved degree of freedom (DoF) and capacity of the proposed approach are analyzed. Simulation results show that, compared to the conventional singular value decomposition (SVD)-based method, the proposed transceiver design can achieve near-optimal DoF and capacity with a significantly lower computational complexity.

\end{abstract}

\begin{IEEEkeywords}
Massive MIMO, transceiver design, channel sparsity, DoF
\end{IEEEkeywords}

\IEEEpeerreviewmaketitle

\section{Introduction}

The multiple-input-multiple-output (MIMO) technique has been known as an effective way to significantly increase the capacity of wireless communications. Theoretically, the capacity of a MIMO system can increase linearly with the minimum number of the transmit (Tx) and receive (Rx) antennas for fixed Tx power and bandwidth \cite{MIMO_Capacity}. When the number of antennas becomes very large, such as in massive MIMO systems, multiple gains (e.g., rate increase and transmission reliability, etc.) can further scale up by possible orders of magnitude compared to the current state-of-art \cite{Massive_MIMO}.

To achieve the capacity of a MIMO system, singular value decomposition (SVD) approach has been widely used in the open literature to decompose the MIMO channel into a set of parallel single-input-single-output (SISO) subchannels, over which multiple independent signal streams can be transmitted without mutual interference \cite{SVD}. However, the computational cost of the SVD-based design becomes prohibitively high in massive MIMO systems. One approach to reduce the computational complexity in such systems is the antenna selection technique, which can achieve a similar diversity gain as the all-participation setup but significantly sacrifice the degree of freedom (DoF) and thereby the capacity of MIMO systems~\cite{Antenna_Selection}.

On the other hand, recent studies have demonstrated that, as the system dimension increases, the physical MIMO channels exhibit a sparse structure due to insufficient scatterers in propagations \cite{MIMO_Sparcity,review1}. Several low-complexity transceivers have been developed in~\cite{Low_Complexity_MIMO_Sparcity} to exploit the channel sparsity of a point-to-point large-scale MIMO system. However, the designs in~\cite{Low_Complexity_MIMO_Sparcity} focused on a particular low-rank millimeter wave scenario, which may fail to capture the full DoF of general multipath massive MIMO systems. To the best of our knowledge, how to capture the full DoF of sparse multipath massive MIMO channels using low-complexity transceiver has not been well addressed in~open~literatures.

Motivated by this, in this letter we develop a novel low-complexity transceiver design, namely semi-random beam pairing (SRBP), for sparse multipath massive MIMO channels. The SRBP algorithm is designed based on the virtual channel model in angular domain \cite{Sayeed_VirtualChannelRepresentation_1,Sayeed_VirtualChannelRepresentation_2}. Specifically, a set of transmit-receive beam pairs are generated in a semi-random way. Each beam pair is used to transmit one data stream such to enable the simultaneous data transmission. These data streams can be easily separated at the receiver via a successive interference cancelation (SIC) technique, and the power allocation among them are optimized based on the classical water-filling principle. An analytical expression for the achieved DoF of SRBP is derived. Numerical results demonstrate that SRBP can achieve near-optimal DoF and capacity performance but has a much lower computational~complexity.


\section{System Model}


Consider a massive MIMO system with an $N_t$-element Tx uniform linear array (ULA) and an $N_r$-element Rx ULA \cite{point_to_point}. By assuming a frequency-flat fading channel between the two ends, the Tx and Rx signals are related by
\begin{equation}\label{Equ:Received_Signal_at_BaseStation}
  {{\bf{y}}} = {{\bf{H}}}{{\bf{x}}} + {\bf{n}},
\end{equation}where $\mathbf{x}\!\in\! \mathcal{C}^{N_t\times1}$ is the Tx signal with $\mathcal{C}^{{m}\times{n}}$ representing the set of all $m\!\times\!n$ complex matrices, $\bf{n}$ is a length-$N_r$ vector of complex additive white Gaussian noise, and $\mathbf{H}\!\in\!{\mathcal{C}^{{N_r}\!\times\!{N_t}}}$ is the physical multipath channel matrix given by~\cite{Sayeed_VirtualChannelRepresentation_1,Sayeed_VirtualChannelRepresentation_2}
\begin{equation}\label{Equ:Physical_Channel_Model}
  {\bf{H}} = \sqrt {{N_r}{N_t}} \sum\nolimits_{\ell = 1}^L {{g_\ell}{{\bf{a}}_r}({\omega _{r,\ell}}){\bf{a}}_t^H({\omega _{t,\ell}})}.
\end{equation}In (\ref{Equ:Physical_Channel_Model}), the Tx and Rx antennas are linked via $L$ propagation paths with complex gains $\{g_\ell\}$, angles of departure (AoDs) $\{{\omega _{t,\ell}}\}$ and angles of arrival (AoAs)  $\{{\omega _{r,\ell}}\}$, and $(\cdot)^H$ represents the conjugate transpose operation. The steering vector ${{\bf{a}}_t}({\omega _{t}})$ and response vector ${{\bf{a}}_r}({\omega _{r}})$ are expressed, respectively, as
\begin{eqnarray}\label{Equ_Tx}
{{\bf{a}}_t}({\omega _t})\!\!\!\! &=&\!\!\!\! \frac{1}{{\sqrt {{N_t}} }}\left[\!\! {\begin{array}{*{20}{c}}
1,\!\!\!&{{e^{ - j2\pi {\omega _t}}}},\!\!\!& \cdots,\!\!\! &{{e^{ - j2\pi {\omega _t}{(N_t\!-\!1)}}}}
\end{array}}\!\! \right]^T,\\
\label{Equ_Rx}
{{\bf{a}}_r}({\omega _r})\!\!\!\! &=&\!\!\!\! \frac{1}{{\sqrt {{N_r}} }}\left[\!\! {\begin{array}{*{20}{c}}
1,\!\!\!&{{e^{ - j2\pi {\omega _r}}}},\!\!\!& \cdots,\!\!\! &{{e^{ - j2\pi {\omega _r}{(N_r\!-\!1)}}}}
\end{array}}\!\! \right]^T,
\end{eqnarray}where $(\cdot)^T$ represents the transpose operation.

As discussed in \cite{Sayeed_VirtualChannelRepresentation_1,Sayeed_VirtualChannelRepresentation_2}, $\bf{H}$ can be characterized and represented by a virtual channel ${\bf{H}}_v$ in angular domain with the following relationship
\begin{eqnarray}\label{Equ:Virtual_MIMO_Channel_Representation}
{\bf{H}}\!\! &=&\!\! \sum\nolimits_{i = 1}^{{N_r}} {\sum\nolimits_{j = 1}^{{N_t}} {{{\bf{H}}_v}(i,j){{\bf{a}}_r}({\bar{\omega} _{r,i}}){\bf{a}}_t^H({\bar{\omega} _{t,j}})} } \nonumber\\
 \!\!&=&\!\! {{\bf{A}}_r}{{\bf{H}}_v}{\bf{A}}_t^H,
\end{eqnarray}where $\{\bar{\omega}_{r,i}\!\!=\!\!i/N_r\}$ and $\{\bar{\omega}_{t,j}\!\!=\!\!j/N_t\}$ are the uniformly sampled \textit{virtual} AoAs and AoDs, respectively, and ${{\bf{H}}_v}(i,j)$ approximately equals to the sum of gains of a sub-set of paths which are unresolvable in the $j$th \textit{virtual} AoD and the $i$th \textit{virtual} AoA. Consequently, ${\bf A}_r$ and ${\bf A}_t$ are discrete Fourier transform matrices, and then the virtual representation ${\bf{H}}_v$ is unitarily equivalent to the physical channel matrix $\bf H$ with the relationship ${\bf H}_v = {\bf A}^H_r {\bf H} {\bf A}_t$.


When the number of antennas increases, resolvable paths that contribute to the channel power gain become less due to the insufficient scatterers in propagations \cite{MIMO_Sparcity,review1}. In other words, there are less non-zero entries in ${\bf{H}}_v$ with other entries being approximate to zero when there are no scattering in the corresponding virtual angles. In this sense, it would be valid to assume that ${\bf{H}}_v$ tends to be sparse. In order to further capture the sparse property, we follow \cite{Sayeed_VirtualChannelRepresentation_1,Sayeed_VirtualChannelRepresentation_2,review2} to express ${\bf{H}}_v$~as

\begin{equation}\label{Equ:Mask_expression}
 {{\bf{H}}_{v}} \approx {{\bf{M}}} \odot {\bf{H}}_{{\rm{iid}}},
\end{equation}where $\odot$ denotes the element-wise product, ${\bf{H}}_{\rm{iid}}$ is an independent and identically distributed (i.i.d.) complex Gaussian matrix, and $\bf M$ is a binary mask matrix with each of its entries being $1$ if its counterpart in ${\bf{H}}_v$ is nonzero and $0$ otherwise. It is worth noting that each non-zero entry ${\bf M}(i,j)$ corresponds to the paths from the $j$th virtual AoD to the $i$th virtual AoA.

For the channel ${\bf{H}}_v$ in (\ref{Equ:Mask_expression}), the channel capacity can be achieved via the optimal SVD-based transceiver design. Mathematically, we have
\begin{eqnarray}\label{Equ_Capacity_SVD}
C(\rho \left| {{{\bf{H}}_v}} \right.)=\!\begin{array}{*{20}{c}}
{\mathop {\max }\limits_{{\rho _{i:}}\sum\nolimits_i {{\rho _{i}} = \rho } } }&{\sum\nolimits_{i = 1}^D {\log (1 + {\rho _i}\lambda _i^2)} },
\end{array}
\end{eqnarray}where $D$ is the rank of ${\bf{H}}_v$ and represents the DoF of systems, $\lambda _i$ is the $i$th singular value of ${\bf{H}}_v$, $\rho$ is the total Tx power and $\rho_i$ is the allocated power based on the optimal water-filling technique for the $i$th eigen channel. However, the computational complexity of the SVD-based method becomes prohibitively high in massive MIMO systems with large-scale antennas. Motivated by this, in this letter we propose a low-complexity transceiver design for the sparse multipath massive MIMO channel elaborated in the next section.

\section{Proposed low-complexity transceiver design}
In this section, a low-complexity transceiver design is developed and analyzed for sparse multipath massive MIMO channels. The proposed design adopts a SRBP approach combined with the water-filling and SIC techniques. For simplicity we elaborate our design based on the symmetric case for $N_t\!\!=\!\!N_r\!\!=\!\!N$. The general case where $N_t\!\!\neq\!\!N_r$ is~ readily~extendable.

\subsection{SRBP-based Transceiver Design}
The main idea of the SRBP is to generate multiple Tx-Rx beam pairs in angular domain, with each pair transmitting one data stream. We aim at generating as many beam pairs as possible so as to approach the full DoF of systems. The SRBP is performed on the binary mask matrix $\bf M$ which maintains the same sparsity as ${\bf{H}}_v$. Specifically, it consists of the following three main~processes.

\begin{figure}
  \centering
  \includegraphics[width=0.55\textwidth]{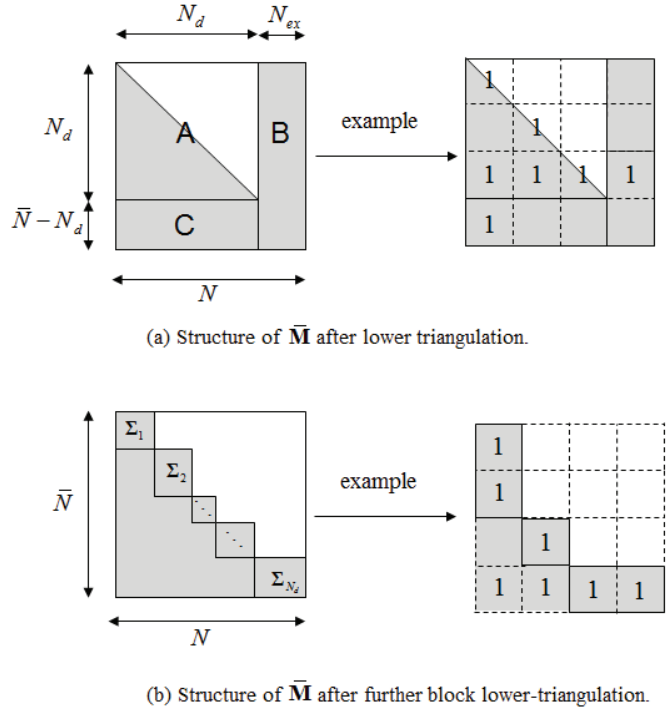}
  \caption{Illustration of ${{\bf{\bar M}}}$ and its example corresponding to Fig.~1.}
  \label{Fig:MatrixOperation}
\end{figure}

\begin{enumerate}
  \item \emph{\textbf{Initialization}}. Remove the all-zero rows in $\bf M$ which make no contribution to channel gains. The downsize matrix of $\bf M$ is denoted by $\bar{\bf M}\in\mathcal{C}^{{\bar{N}}\times{N}}$.
  \item \emph{\textbf{Lower-triangulation}}. This process typically consists of $N$ steps. At the $\ell$th step, the following operations are performed on the operating matrix $\bar{\bf M}^{(\ell)}$ (which is a sub-matrix of $\bar{\bf M}$) as illustrated in blue color in Fig.~\ref{Fig:Iteration}.

  \begin{figure}
  \centering
  \includegraphics[width=0.7\textwidth]{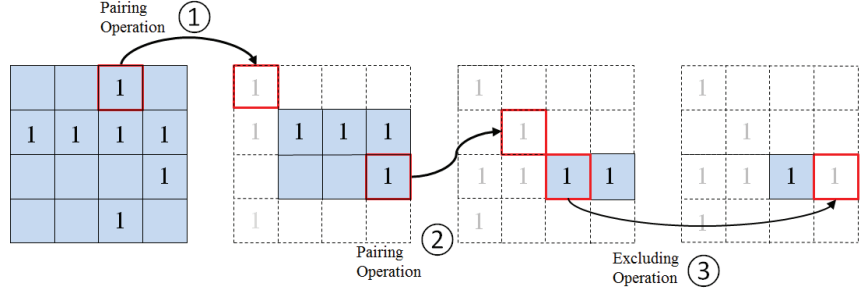}
  \caption{{Illustration of the pairing and temporarily excluding procedures.}}
  \label{Fig:Iteration}
\end{figure}

\begin{enumerate}
\item \textit{Beam pairing}:
      \begin{itemize}
        \item Find a weight-$1$ row\footnote{We hereafter term a row with $k$ non-zero entries as a weight-$k$ row.} in $\bar{\bf M}^{(\ell)}$ and move this unique ``$1$" \big(e.g., in the $i$th row and $j$th column) to the top-left of $\bar{\bf M}^{(\ell)}$ via row/column permutations to obtain $\bar{\bf{M}}_{temp}^{\left( \ell  \right)}$. In mathematics,
            \begin{eqnarray}\label{M_beam_pairing}
            {{{\bar{\bf{M}}_{temp}}}^{\left( \ell  \right)}}\!=\!{\bf{\bar{E}}}_{i}^{\left( \ell  \right)^T}{{\bf{\bar M}}^{\left( \ell  \right)}}{\bf{\bar{E}}}_j^{\left( \ell  \right)},
            \end{eqnarray}where ${\bf{\bar{E}}}_i^{(\ell )}\!\!=\!\! \left[ {{{\bf{e}}_i},{{\bf{e}}_1}, \cdots ,{{\bf{e}}_{i - 1}},{{\bf{e}}_{i + 1}}, \cdots {{\bf{e}}_{{m^{(\ell )}}}}} \right]$ is the row permutation matrix, $m^{(\ell)}$ is the row dimension of $\bar{\bf M}^{(\ell)}$ and ${\bf{e}}_i$ is the unit vector with $1$ at the $i$th entry and zeros otherwise. The column permutation matrix ${\bf{\bar{E}}}_j^{\left( \ell  \right)}$ can be obtained~similarly.
        \item Update $\bar{{\bf M}}_{temp}^{(\ell)}$ to $\bar{\bf M}^{(\ell+1)}$ by removing the original $i$th row, $j$th column and the resulting all-zero~rows.
      \end{itemize}
  \item \textit{Temporarily column excluding}:
      \begin{itemize}
    \item If there is no weight-$1$ row in $\bar{\bf M}^{(\ell)}$, we random select one (e.g., $k$th) column in $\bar{\bf M}^{(\ell)}$ and move it to the right most of $\bar{\bf M}^{(\ell)}$ to obtain $\bar{\bf{M}}_{temp}^{\left( \ell  \right)}$. In~mathematics,
         \begin{eqnarray}\label{M_random_excluding}
            {{{\bar{\bf{M}}_{temp}}}^{\left( \ell  \right)}}\!=\!{{\bf{\bar M}}^{\left( \ell  \right)}}{\bf{\underline{E}}}_k^{\left( \ell  \right)},
            \end{eqnarray}where ${\bf{\underline{E}}}_k^{(\ell )}\!\!=\!\!\left[ {{{\bf{e}}_1}, \cdots ,{{\bf{e}}_{k\!-\!1}},{{\bf{e}}_{k\!+\!1}}, \cdots {{\bf{e}}_{N\!-\!\ell\!+\!1}},{{\bf{e}}_k}} \right]$.
    \item Update $\bar{{\bf M}}_{temp}^{(\ell)}$ to $\bar{\bf M}^{(\ell+1)}$ by removing the original $k$th column and the resulting all-zero rows.

      \end{itemize}
\end{enumerate}

After this process, the matrix $\bar{\bf M}$ will be permuted into the form as shown in Fig.~\ref{Fig:MatrixOperation}(a), where ${\bf A}\!\in\!{\mathcal C}^{N_d\!\times\! N_d}$ is a lower triangular matrix and ${\bf B} \!\in\! {\mathcal C}^{{\bar N}\! \times\! N_{ex}}$ consists all the temporarily excluded columns. Specifically, $N_d$ denotes the achieved DoF and $N_{ex}$ denotes the times of temporarily excluding operations.

\item \emph{\textbf{Further block lower-triangulation}}. This process aims at utilizing the non-zero entries in $\bf B$ and $\bf C$ to achieve the potential power gains.

Scan the rows of $\bf B$ from top to down. Assuming there are $q$ ($q\!\leq\!N_{ex}$) non-zero entries in the $i$th ($i\!\leq\!N_d$) row of $\bf B$ for example. Move these $q$ $1$'s via column permutations to right next to the diagonal $1$ in the $i$th row of $\bf A$. Similarly, scan the columns of $\bf C$ from right to left and move all $1$'s in $\bf C$ upwards to next to the diagonal blocks in $\bf A$ via row permutations.

With this process, as showed in Fig.~\ref{Fig:MatrixOperation}(b), the small blocks $\{{\bf \Sigma}_i\}$ on the diagonal are created by the corresponding non-zero entries in $\bf B$ and $\bf C$ together with the diagonal elements in $\bf A$.
\end{enumerate}

Finally, the corresponding complex channel matrix $\tilde{{\bf H}}_v$ can be mapped from $\bar{\bf M}$, and the diagonal blocks in $\tilde{{\bf H}}_v$ is denoted by $\{\tilde{{\bf \Sigma}}_i\}$. The eigen channel corresponding to the largest singular value of each $\tilde{{\bf \Sigma}}_i$ then transmits one data stream. Moreover, by representing $\tilde{{\bf H}}_v$ in the block lower-triangular form, the receiver can adopt the SIC technique to cancel the interferences among multiple data streams. In this sense, the separable data streams can be treated as parallel at the transmitter. The classical water-filling principle is thus adopted to achieve the optimal power allocation among all data streams. Mathematically, the achieved capacity of the proposed transceiver design can be expressed as
\begin{eqnarray}\label{Equ_Capacity_SRBP}
C(\tilde \rho \left| {{{{\tilde{\bf H}}}_v}} \right.) = \begin{array}{*{20}{c}}
{\mathop {\max }\limits_{{{\tilde \rho }_{i:}}\sum\nolimits_i {{{\tilde \rho }_{i}} = \rho } } }&{\sum\nolimits_{i = 1}^{{N_d}} {\log (1 + {{\tilde \rho }_i}\tilde \lambda _i^2)} },
\end{array}
\end{eqnarray}
where ${\tilde \lambda_i}$ is the largest singular value of ${\tilde{\bf{\Sigma}}}_i$.

\emph{\textbf{Remark of Complexity}}: In the proposed SRBP transceiver design, the computational complexity is mainly contributed by two aspects. The first aspect is the generation of the diagonal blocks $\{{{\tilde{\bf \Sigma}}}_i\}$, which only involves row/column permutations with very low complexity. The second one is the SVD of $\{{{\tilde{\bf \Sigma}}}_i\}$ to obtain the maximum eigen values $\{\tilde \lambda_i\}$.

It is worth noting that the maximum size of ${{\tilde{\bf \Sigma}}}_i$ is upper-bounded by $(\bar{N}\!-\!N_d\!+\!1)\!\times\!(N_{ex}\!+\!1)$, which is much smaller than that of ${{\bf H}}_v$ for $N\!\times\!N$. In this case, the SVD of $\{{\tilde{\bf \Sigma}}_i\}$ has much lower complexity given by $O\big((\bar{N}\!-\!N_d\!+\!1)(N_{ex}\!+\!1)^2\big)$ than that of the full-size channel matrix ${\bf{H}}_v$ for $O(N^3)$ \cite{SVD_Complexity}.

Furthermore, from the statistical perspective, the probability that a diagonal block ${{\tilde{\bf \Sigma}}}_i$ with the exact size $(\bar{N}\!-\!N_d\!+\!1)\!\times\!(N_{ex}\!+\!1)$ in $\tilde{{\bf H}}_v$ is extremely small, as it requires the entries in a certain row of $\bf B$ and a corresponding column of $\bf C$ are all non-zeros. This deduction is actually verified in the latter numerical results, in which we show that more than $99\%$ of $\{{{\tilde{\bf \Sigma}}}_i\}$ are either single element or vector. We thus can claim that the actual complexity of SRBP is further lower than the aforementioned upper bound.

\subsection{DoF Analysis of the SRBP Algorithm}
In this subsection we derive an analytical expression for the average achieved DoF of the proposed SRBP algorithm.

As in \cite{Gaussian_distribution}, we adopt the Bernoulli distribution to model the channel sparsity. Specifically, each entry of $\bf M$ is assumed i.i.d. and to take value $1$ with a small probability $\delta$ and value $0$ with the probability $1\!-\!\delta$. Let $\hat{p}_k$ denote the probability that a row of ${\bf M}$ has $k$ non-zero entries.
 When the number of antennas $N$ goes infinite, $\hat{p}_k$ follows a Poisson distribution with the probability mass function \cite{Poisson_distribution}
\begin{eqnarray}\label{Equ:Original_row_degree}
{\hat{p}_{k}} = {{{e^{ - {\beta}}}\times\beta^k}}/{{k!}},~~k\!=\!\{ 0,1,\cdots,N\},
\end{eqnarray}where ${\beta}\!=\!N\!\times\!\delta$ denotes the average number of $1$'s in each~row.

After the \textit{initialization} process, ${\bf M}$ is down-sized to $\bar{\bf M}^{(1)}$ with ${m}^{(1)}$ rows on average, which is associated to $N$ and $\hat{p}_0$~as
\begin{eqnarray}\label{Equ:Nomalized_row_number}
{{m}^{(1)}} \!=\! N \times (1-\hat{p}_0),
\end{eqnarray}and the probability that a row in $\bar{\bf M}^{(1)}$ has $k$ non-zero entries is updated by
\begin{eqnarray}\label{Equ:Nomalized_row_degree}
p_{k}^{(1)} = {\hat{p}_{k}}/(1-\hat{p}_0),\quad k = \{ 1, \cdots ,{N}\}.
\end{eqnarray}In (\ref{Equ:Nomalized_row_number}) and (\ref{Equ:Nomalized_row_degree}), the superscript $(\cdot)^{(1)}$ denotes the value of a certain parameter after initialization and before the first step of \emph{lower-triangulation process}.

Note that after the $N$ columns are either paired or temporarily excluded in the \emph{lower-triangulation process}, as illustrated in Fig.~\ref{Fig:MatrixOperation}, the relationship between the average achieved DoF ($N_d$) and the average times of temporarily excluding operations ($N_{ex}$) is given~by
\begin{eqnarray}\label{Equ_N_d}
{N_d} = {N} - {N_{ex}}.
\end{eqnarray}
In this case, the analytical expression of $N_d$ will be obtained if we can find the expression of $N_{ex}$. To proceed, we denote by $p_{ex}^{(\ell)}$ the probability that the temporarily excluding operation occurs in the $\ell$th step. Then $N_{ex}$ can be obtained by summing $p_{ex}^{(\ell)}$ over $\ell$ for $\ell\in\{1,\ldots,N\}$, i.e.,
\begin{eqnarray}\label{Equ:Inactivation_total}
{N_{ex}} = \sum\nolimits_{\ell = 1}^{N} {p_{ex}^{(\ell)}}.
\end{eqnarray}
With the values of ${{m}^{(1)}}$ and $p_{k}^{(1)}$, we first can obtain
\begin{equation}\label{eq:p_ex_1}
{p_{ex}^{(1)}} = {(1 - p_{1}^{(1)})^{{{m}^{(1)}}}}
\end{equation}
by realizing that the excluding operation occurs only when there is no weight-$1$ row in the current operating matrix $\bar{\bf M}^{(1)}$. The value of ${p_{ex}^{(\ell)}}$ for $\ell\in\{2,\ldots,N\}$ should be calculated sequentially as the execution of \emph{lower-triangulation process} step by step. Without loss of generality, in the following we explain how to calculate ${p_{ex}^{(\ell)}}$ for the $\ell$th step based on the parameters obtained in the $(\ell\!-\!1)$th step. That is, all notations with superscript $(\cdot)^{(\ell\!-\!1)}$ have already been known. Besides, all other involved notations in this calculation and their corresponding physical meanings are listed in Table. \ref{tab:table0}.

\begin{table}
\begin{center}
\caption{List of notations and their physical meanings}\label{tab:table0}
\begin{tabular}{|c|l|}
\hline
\footnotesize Notation &  Physical Meaning \\
\hline
\footnotesize $\bar{\mathbf{M}}^{(\ell)}$ &  Operating Matrix in the $\ell$th step, $\ell\in\{1,\ldots,N\}$.  \\
\footnotesize $m^{(\ell)}$ & The average row-dimension of matrix $\bar{\mathbf{M}}^{(\ell)}$.  \\
\footnotesize $N_k^{(\ell)}$ & The average number of weight-$k$ rows in $\bar{\mathbf{M}}^{(\ell)}$ \\
\footnotesize  &for $k\in\{1,\ldots,N-(\ell\!-\!1)\}$.\\
\footnotesize $p_k^{(\ell)}$ & The probability that a row of $\bar{\mathbf{M}}^{(\ell)}$ is weight-$k$.   \\
\footnotesize $Q_k^{(\ell)}$ & The average number of rows reducing weight from $k$ to ($k\!-\!1$) \\
\footnotesize  &  when updating $\bar{\mathbf{M}}^{(\ell)}$ in the $\ell$th step. \\
\footnotesize $\alpha_k^{(\ell)}$ & The probability that a row reducing weight from $k$ to ($k\!-\!1$)\\
\footnotesize  &  when updating $\bar{\mathbf{M}}^{(\ell)}$ in the $\ell$th step. \\
\hline
\end{tabular}
\end{center}
\end{table}

Similar to (\ref{eq:p_ex_1}), $p_{ex}^{(\ell)}$ can be expressed~as
\begin{eqnarray}\label{Equ:Inactivation_each_step}
{p_{ex}^{(\ell)}}\!
\! =\!{(1 - p_{1}^{(\ell)})^{{{m}^{(\ell)}}}}.
\end{eqnarray}
We thus need to calculate $p_{1}^{(\ell)}$ and ${{{m}^{(\ell)}}}$ based on the parameters obtained in $(\ell\!-\!1)$th step.

Recall that in each step of \emph{lower-triangulation process}, one column is removed from the operating matrix no matter \emph{beam pairing} or \emph{temporarily column excluding} procedure is performed. In this sense, the weight (i.e., number of $1$'s) of each row may be reduced when updating the operating matrix. We assume there are a number of $Q_1^{(\ell-1)}$ rows on average changed from weight-$1$ to all-zeros that should be removed when updating $\bar{\mathbf{M}}^{(\ell\!-\!1)}$ to $\bar{\mathbf{M}}^{(\ell)}$. 
Thus ${m}^{(\ell)}$ can be related with ${m}^{(\ell-1)}$ as
\begin{eqnarray}\label{Equ:Updated_m}
{m}^{(\ell)}={m}^{(\ell\!-\!1)}\!-\! Q_1^{(\ell\!-\!1)},~~\ell\in\{2,\ldots,N\}.
\end{eqnarray}

The value of the term $Q_1^{(\ell\!-\!1)}$ in (\ref{Equ:Updated_m}) is contributed by two parts. The first part comes from the selected weight-$1$ row of $\bar{\mathbf{M}}^{(\ell\!-\!1)}$ in \emph{Beam pairing} procedure, which becomes an all-zero row in $\bar{\mathbf{M}}^{(\ell\!)}$ with probability $1$. The second part is from the remaining $(N_1^{(\ell\!-\!1)}\!-\!1)$ weight-$1$ rows in $\bar{\mathbf{M}}^{(\ell\!-\!1)}$, which may also be changed to weight-$0$ when removing the column containing the selected ``1". Since the positions of 1's in different rows are considered to be independent, we can assume that the remaining $(N_1^{(\ell\!-\!1)}\!-\!1)$ weight-$1$ rows reduce weight with the same probability $\alpha _1^{(\ell\!-\!1)}$. Considering that the value of $N_1^{(\ell\!-\!1)}$ could range from $1$ to ${m}^{(\ell\!-\!1)}$ and according to the full probability theory, $Q_1^{(\ell\!-\!1)}$ can be given~by
\begin{eqnarray}\label{Equ:Number_Degree_1_rows_deleted}
\small
Q_1^{(\ell\!-\!1)}\!\!\!\!\!&=&\!\!\!\!\!\sum\limits_{t = 1}^{{{m}^{(\ell\!-\!1)}}}\bigg\{ \Pr(N_1^{(\ell\!-\!1)}=t) {\left[ {1 \!+\! (t \!- \!1)  \alpha_1^{(\ell\!-\!1)}} \right]} \bigg\}\nonumber\\
\!\!\!\!\!&=&\!\!\!\!\! (1\! -\!\alpha_1^{(\ell\!-\!1)})\!\!\sum\limits_{t \!=\! 1}^{{{m}^{(\ell\!-\!1)}}}\Pr(N_1^{(\ell\!-\!1)}=t) \!+\! \alpha_1^{(\ell\!-\!1)} \mathbb E[N_1^{(\ell\!-\!1)}]\nonumber\\
\!\!\!\!\!&=&\!\!\!\!\! (1\!-\! \alpha_1^{(\ell\!-\!1)})(1 \!-\! p_{ex}^{(\ell\!-\!1)})\!+\! \alpha_1^{(\ell\!-\!1)}{m}^{(\ell\!-\!1)}p_{1}^{(\ell\!-\!1)},
\end{eqnarray}
where $\mathbb E[\cdot]$ represents the expectation of a random variable. Recall that the position of non-zero entries in each row are assumed to be randomly distributed, then the more non-zero entries a row has, the larger the probability that the weight will be reduced when removing column in each step of \emph{lower-triangulation process}. In this sense, $\alpha _1^{(\ell\!-\!1)}$ can be given~by
\begin{eqnarray}\label{Equ:Probability_Degree_1_affected}
\alpha _1^{(\ell\!-\!1)}\!\! =\!\! {1}/{{\left[N\!-\!(\ell\!-\!1)\!+\!1\right]}}.
\end{eqnarray}
as $\bar{\mathbf{M}}^{(\ell\!-\!1)}$ has  ${\left[N\!-\!(\ell\!-\!1)\!+\!1\right]}$ entries in each row.

So far, ${m}^{(\ell)}$ can be obtained by substituting (\ref{Equ:Probability_Degree_1_affected}) into (\ref{Equ:Number_Degree_1_rows_deleted}) and then (\ref{Equ:Number_Degree_1_rows_deleted}) into (\ref{Equ:Updated_m}). We now turn to the calculation of $p_1^{(\ell)}$, which is defined as the ratio of ${N}_1^{(\ell)}$ and ${m}^{(\ell)}$. Mathematically,
\begin{eqnarray}\label{Equ:Probability_i_updated}
p_1^{(\ell)} = {{{N}_1^{(\ell)}}}/{{{{m}^{(\ell)}}}},
\end{eqnarray}
where the value of ${{{N}_1^{(\ell)}}}$ can be derived from ${{{N}_1^{(\ell-1)}}}$ by first  subtracting the average number of rows reduced from weight-$1$ to weight-$0$ (i.e., $Q_1^{(\ell-1)}$) and then adding the average number of rows reduced from weight-$2$ to weight-$1$ (i.e., $Q_2^{(\ell-1)}$) when updating $\bar{\mathbf{M}}^{(\ell\!-\!1)}$ to $\bar{\mathbf{M}}^{(\ell)}$. That~is,
\begin{eqnarray}\label{Equ:N_k}
N_1^{(\ell)}=N_1^{(\ell\!-\!1)} - {Q}_1^{(\ell\!-\!1)} + {Q}_{2}^{(\ell\!-\!1)},
\end{eqnarray}
where ${Q}_2^{(\ell\!-\!1)}$ is readily given by
\begin{eqnarray}\label{Equ:Degree_i_affected}
{Q}_2^{(\ell\!-\!1)} = \alpha _2^{(\ell\!-\!1)}{N}_2^{(\ell\!-\!1)}.
\end{eqnarray}Similar to (\ref{Equ:Probability_Degree_1_affected}), $\alpha _2^{(\ell\!-\!1)}$ can be expressed by
\begin{eqnarray}\label{Equ:Probability_Degree_affected}
\alpha _2^{(\ell\!-\!1)}\!\! =\!\! {2}/[{{N\!-\!(\ell\!-\!1)+1}}].
\end{eqnarray}Finally, by substituted (\ref{Equ:Probability_Degree_affected}) into (\ref{Equ:Degree_i_affected}) then (\ref{Equ:N_k}) into (\ref{Equ:Probability_i_updated}), the value of $p_1^{(\ell)}$ in (\ref{Equ:Inactivation_each_step}) have been achieved. At the same time, we complete the calculation of ${p_{ex}^{(\ell)}}$. Note that due to the inherent iterative feature of the proposed SRBP scheme, it is difficult to write a close-form expression for its achievable DoF. But it can be numerically calculated in an iterative way as elaborated in (\ref{Equ_N_d})-(\ref{Equ:Probability_Degree_affected}).

\section{Numerical Results}
We now provide some numerical results to illustrate the SRBP performance. In the following simulations, we follow the typical multiplexing configuration in \cite{Sayeed_VirtualChannelRepresentation_2} and set $\delta\!=\!1/N$.

Table.~II verifies the theoretical analysis of the achieved DoF of the sparse multipath massive MIMO. We can see that the analytical and simulation results of SRBP are very close to each other, which validates our theoretical analysis. Meanwhile, compared to the optimal SVD-based scheme, the proposed SRBP design can achieve nearly the same DoF for various $N$.

\begin{table}
\begin{center}
\caption{Achieved DoF comparison}\label{tab:table1}
\begin{tabular}{|c|c|c|c|c|c|}
\hline
\footnotesize Number of antennas &  \footnotesize 8 &  \footnotesize 16 &  \footnotesize 32 & \footnotesize 64 & \footnotesize 128 \\
\hline
\footnotesize DoF: SRBP, analytical, Eq. (\ref{Equ_N_d}) & \footnotesize 4.59 & \footnotesize 8.95 & \footnotesize 17.56 & \footnotesize 34.8 & \footnotesize 69.99 \\
\hline
\footnotesize DoF: SRBP, simulation & \footnotesize 4.43 & \footnotesize 8.74 & \footnotesize 17.39 & \footnotesize 34.59 & \footnotesize 69.83 \\
\hline
\footnotesize DoF: SVD, simulation & \footnotesize 4.49 & \footnotesize 8.79 & \footnotesize 17.44 & \footnotesize 34.64 & \footnotesize 69.88 \\
\hline
\end{tabular}
\end{center}
\end{table}

\begin{figure}
  \centering
  \includegraphics[width=0.6\textwidth]{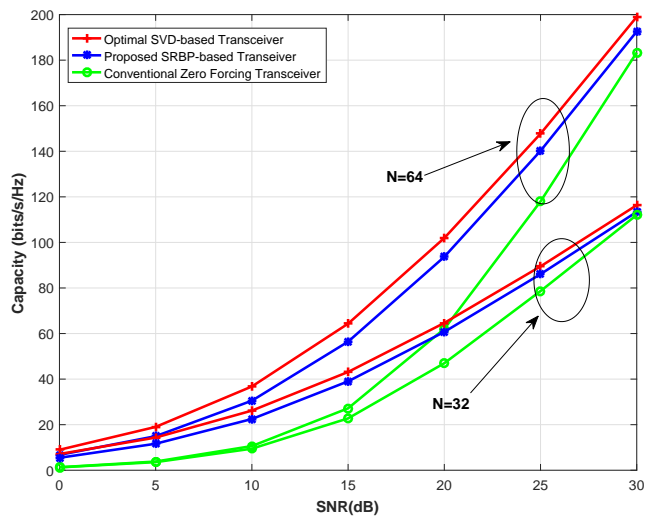}
  \caption{Achievable capacity versus SNR, $N\!\!=\!\!32$ and $N\!\!=\!\!64$.}
  \label{Fig:ZFWF}
\end{figure}

Fig.~\ref{Fig:ZFWF} shows the average achievable capacity versus SNR for $N\!\!=\!\!32$ and $N\!\!=\!\!64$. We can observe in this figure that the capacity of the proposed SRBP-based transceiver design can approach that of the optimal SVD-based transceiver design over the entire SNR region. Note that the simulation results of \cite{Low_Complexity_MIMO_Sparcity} are not compared in this letter as they considered a particular low-rank scenario and did not utilize the full DoF of the system. Hence it is not suitable to solve the considered problem in this paper.

Next, we numerically compare the computational complexity of the proposed transceiver design and the conventional SVD-based one. For $N\!=\!64$, we find that the average number of $N_{ex}\!\approx\!5.8$ and $(\!\bar N\!-\!N_d)\!\approx\!5.4$ over $10,000$ random channel realizations. As showed in Table. \ref{tab:table2}, more than $99\%$ of $\{{\tilde{\bf \Sigma}}_i\}$ are actually either single elements or vectors. This observation confirms the remark that the proposed SRBP transceiver design has much lower computational complexity than that of the conventional SVD-based one.
\begin{table}
\begin{center}
\caption{The size classification of $\{{\tilde{\bf \Sigma}}_i\}$}\label{tab:table2}
\begin{tabular}{|c|c|c|}
\hline
\footnotesize Size of $\{{\tilde{\bf \Sigma}}_i\}$ &  \footnotesize Average Number &  \footnotesize Percentage  \\
\hline
\footnotesize Single entry, $1\!\times\!1$ & \footnotesize 24.65 & \footnotesize $70.93\%$  \\
\hline
\footnotesize Row vector, $a\!\times\!1, a\!\leq\!N_{ex}\!+\!1$ & \footnotesize 4.76 & \footnotesize $14.03\%$  \\
\hline
\footnotesize column vector, $1\!\times\!b, b\!\leq\!(\bar N\!-\!N_d)\!+\!1$ & \footnotesize 5.17 & \footnotesize $14.95\%$  \\
\hline
\footnotesize others & \footnotesize $0.42$ & \footnotesize $0.09\%$  \\
\hline
\end{tabular}
\end{center}
\end{table}
\section{Conclusion}
A low-complexity transceiver design has been proposed and analyzed for sparse multipath massive MIMO channels. Compared to the optimal SVD-based approach over the full-dimension channel matrix, the proposed SRBP method can capture nearly the same DoF  and capacity of MIMO systems while significantly reducing the computation complexity. 



\ifCLASSOPTIONcaptionsoff
  \newpage
\fi


\begin{thebibliography}{1}

\bibitem{MIMO_Capacity}
T.~L.~Mazetta and B.~M.~Hochwald, ``Capacity of a mobile multiple-antenna communication link in Rayleigh flat fading,''~\emph{IEEE Trans. Inf. Thoery}, vol.~45, no.~1, pp.~139-157, Jan. 1999.

\bibitem{Massive_MIMO}
E.~G.~Larsson, O.~Edfors, F.~Tufvesson, and T.~L.~Marzetta, ``Massive MIMO for next generation wireless systems,''\emph{IEEE Comm. Mag.,} vol.~52, pp.~186-195, Feb. 2014.

\bibitem{SVD}
D.~Tse, and P.~Viswanath, \emph{Fundamentals of wireless communication}, Cambridge university press, 2005.

\bibitem{Antenna_Selection}
A.~F.~Molisch, ``MIMO systems with antenna selection - an overview," \emph{IEEE Radio and wireless conference (RAWCON),} pp. 167-170, Aug. 2003.

\bibitem{MIMO_Sparcity}
M.~Michail, A.~M.~Sayeed, and J.~Nossek, ``Sparse multipath MIMO channels: Performance implications based on measurement data," in 10$^{th}$ Workshop on \emph{IEEE Signal Process.~Advances in Wireless Commun. (SPAWC)}, Perugia, pp. 364-368, Jun. 2009.

\bibitem{review1}
G.~Zhen, et al., ``Super-resolution sparse MIMO-OFDM channel estimation based on spatial and temporal correlations,"~\emph{IEEE Commun.~Lett.}, vol. 18, pp.~1266-1269, May 2014.

\bibitem{Low_Complexity_MIMO_Sparcity}
G.~H.~Song,~J.~Brady, and A.~M.~Sayeed, ``Beamspace MIMO transceivers for low-complexity and near-optimal communication at mm-wave frequencies," in \emph{IEEE ICASSP}. pp. 4394-4398, 2013.

\bibitem{Sayeed_VirtualChannelRepresentation_1}
A.~M.~Sayeed, ``Deconstructing multi-antenna fading channels,''~\emph{IEEE Trans. Signal Process.}, vol. 50, no. 10, pp. 2563-2579, Oct. 2002.

\bibitem{Sayeed_VirtualChannelRepresentation_2}
A.~M.~Sayeed and V.~Raghavan, ``Maximizing MIMO capacity in sparse multipath with reconfigurable antenna arrays,''~\emph{IEEE J. Sel. Topics Signal Process.}, vol.~1, no.~1, pp.~156-166, Jun. 2007.

\bibitem{point_to_point}
J.~Chen, and V.~K.~Lau, ``Multi-stream iterative SVD for massive MIMO communication systems under time varying channels," in~\emph{IEEE ICASSP}, pp. 3125-3156, 2014.

\bibitem{review2}
G,~Zhen, et al. ``Spatially common sparsity based adaptive channel estimation and feedback for FDD massive MIMO,"~\emph{IEEE Trans. Signal Process.}, vol.~63, pp.~6169-6183, Dec. 2015.

\bibitem{SVD_Complexity}
L.~N.~Trefethen, D.~B.~III, \emph{Numerical linear algebra,} Society for industrial and applied mathematics (SIAM), 1997.

\bibitem{Gaussian_distribution}
L.~Cai, P.~Wang, Y.~Li, et al., ``Asymptotic capacity analysis for sparse multipath multiple-input multiple-mutput mhannels,''~\emph{IEEE Commun. Lett.}, vol. 19, no. 12, pp.~2262-2265, Dec. 2015.

\bibitem{Poisson_distribution}
NIST/SEMATECH,~\emph{e-Handbook~of~Statistical~Methods},\\
http://www.itl.nist.gov/div898/handbook/.

\end{thebibliography}
\end{document}